\documentclass[11pt]{article}
\usepackage{paspconf}
\usepackage{graphics}
\begin{document}
\vglue -2cm
\noindent {\sl Contribution to Workshop on Cosmic Flows, held in
Victoria, BC, CANADA, July 13--17, 1999, ed. S. Courteau, M. Strauss \&
J. Willick, ASP series}

\vspace{1cm}

\title{Near-Infrared Galaxy Surveys in 2D, 3D \& 4D}

\author{G. A. Mamon\altaffilmark{1}}
\affil{IAP, F-75014 Paris, FRANCE}


\altaffiltext{1}{also DAEC, Observatoire de Paris, F-92195 Meudon, FRANCE}



\begin{abstract}
The completeness and reliability of the {\sf DENIS} $IJK$ survey and the
{\sf EDSGC} (derived from the
{\sf COSMOS} scans of {\sf USKT} plates) are obtained by detailed
cross-identifications 
and systematic visual inspections of conflictual classifications. 
The {\sf DENIS} galaxy extraction turns out to be over 95\% complete and
reliable 
out to $I < 16$, while
the {\sf COSMOS/EDSGC} galaxy catalogue is less than 80\% complete and
reliable at 
$B_J = 17.5$, and even less than 10\% complete at $B_J < 14.5$.

Spectroscopic followups of {\sf DENIS} and the similar Near-IR {\sf 2MASS}
survey are 
described: 1) a redshift survey of 120,000 galaxies using the {\sf 6dF}
robotic multi-fiber spectrograph, 
currently under construction at the {\sf UKST}, and for which a total of 300
nights are guaranteed for 2001-2003, 2) a peculiar velocity survey of
12,000 early-type galaxies with the {\sf 6dF}, and 3) the {\sf DENIS-HI}
peculiar velocity survey of 
5000 inclined spirals visible from Nan\c{c}ay  ($\delta > -38^\circ$), which
has just begun.
The {\sf DENIS-HI} and {\sf 6dF} peculiar velocity samples will have the
strong advantage 
of covering entire regions of the southern sky, and combined, will
multiply by 10 and 4 respectively
the projected and space number densities of objects in the Southern sky.
These two surveys should thus
provide considerably more accurate estimates of the bulk flow, 
$\Omega_{\rm matter}^{0.6}/$bias,
$\Omega_{\rm matter}$ itself, 
and the primordial density fluctuation spectrum.

\end{abstract}




\section{Near-infrared surveys and cosmology}

As is well known, near-infrared (NIR) photons are up to 10 times less
affected by dust extinction than their visible counterparts.
Hence, cosmological surveys based upon NIR-selected galaxy samples can probe
the Universe almost completely through the Zone of Avoidance in the 
Galactic Plane, and moreover have a clearer view of galaxies, virtually
unobscured by dust (especially important for the central regions of spirals and
for edge-on spirals).

The other main advantage of NIR-selected galaxy samples is that they are not
biased towards recent star formation, contrary to optically-selected samples
(especially in blue light), and hence the NIR-selected
galaxies are more weighed by their
mass, or at least their stellar mass.

However, the {\sf IRAS} galaxy samples, heavily weighted towards
galaxies with recent star formation, appear to be better tracers of the
mass in the Universe than are early-type galaxy samples.
The first spectroscopic followups of
NIR selected galaxy samples show fairly small fractions of galaxies
with emission lines in their spectra:
37\% in $J < 13.7$ samples (Mamon et al. 1999)
and 30\% in $K < 12.2$ samples (Huchra, 1999), in comparison with 60\% in
similar depth $B$-selected samples.
Hence, {\sf DENIS} and {\sf 2MASS} will be more weighted towards early-type
galaxies and 
may lead to stronger biases relative to the dark matter
than the {\sf IRAS} samples, but this remains to be seen (especially because,
contrary to the {\sf IRAS} samples, they
will have a mix of early- and late-type galaxies). 

\section{{\sf DENIS}}

The {\sf DENIS} (DEep Near Infrared Southern Sky Survey) consortium is
undertaking 
a complete imaging survey of the southern sky ($-88^\circ < \delta <
+2^\circ$) in the 
Gunn $I \,(0.8\,\mu)$,
Johnson $J \,(1.25\,\mu$), and
$K_s \,(2.15\,\mu)$ bands (Epchtein et al. 1997), and at this writing, 
62\% of the southern sky has been observed
(see {\sf http://www.iap.fr/users/gam/DENIS/slots.html}). The completion date
is planned for January 2001.
The {\sf DENIS} team has built an IR camera and mounted it on the ESO 1m
telescope 
(which had previously only done optical aperture photometry).
The $I$, $J$, and $K$ images are obtained simultaneously through the use of
dichroics with integration times of 9 sec.
The pixel sizes are $1''$ in $I$ and $3''$ in $J$ and $K$, and the latter two
are dithered in 9 sub-exposures of 1 sec each to yield images with
a pseudo-resolution of $1''$.

Star/galaxy separation is performed with classical estimators, and except in
the Zone of Avoidance ($|b| < 2^\circ$), the much higher sensitivity and
spatial resolution of the $I$ band allow us to base our star/galaxy
separation on the $I$ images for objects detected in the $J$ or $K$ images.

\section{Comparison of {\sf DENIS} and {\sf COSMOS/EDSGC} galaxy extractions}

We have performed detailed cross-identifications of the {\sf DENIS}
extractions with 
the {\sf COSMOS} extractions from $B_J$ photographic plates (obtained over the
World Wide Web), over 8 {\sf DENIS} strips,
corresponding to $50\, \rm deg^2$. These strips were randomly selected  at
typically high galactic latitudes $(\langle |b| \rangle \simeq 50^\circ$)
from the
great majority of strips that are not flagged as poor quality.

To compare the quality of the two surveys, we performed two
cross-identifica\-tions: 1) relative to all {\sf DENIS} objects; and 2)
relative to 
{\sf COSMOS} objects ($B_J < 20.5$) in the exact geometry of the 8 {\sf
DENIS} strips. 
We assumed that all objects, on which both {\sf DENIS} and {\sf COSMOS}
agreed were 
galaxies, were indeed galaxies.
We systematically inspected visually the {\sf DENIS} $I$-band images for all
cases 
with $I < 16$ (our estimated limit for 95\% reliable visual classification)
where one of the surveys identified a galaxy while the other called it a star
or did not detect it altogether.
Objects that both {\sf DENIS} and {\sf COSMOS} classify as galaxies have
$\langle 
B-I\rangle = 2.3$ and 95\% have $B-I > 1.5$. Hence, our estimates of {\sf
COSMOS} 
completeness are limited to $B < 17.5$. 

\begin{figure}[ht]
\begin{center}
{\resizebox{!}{0.49\hsize}{\includegraphics{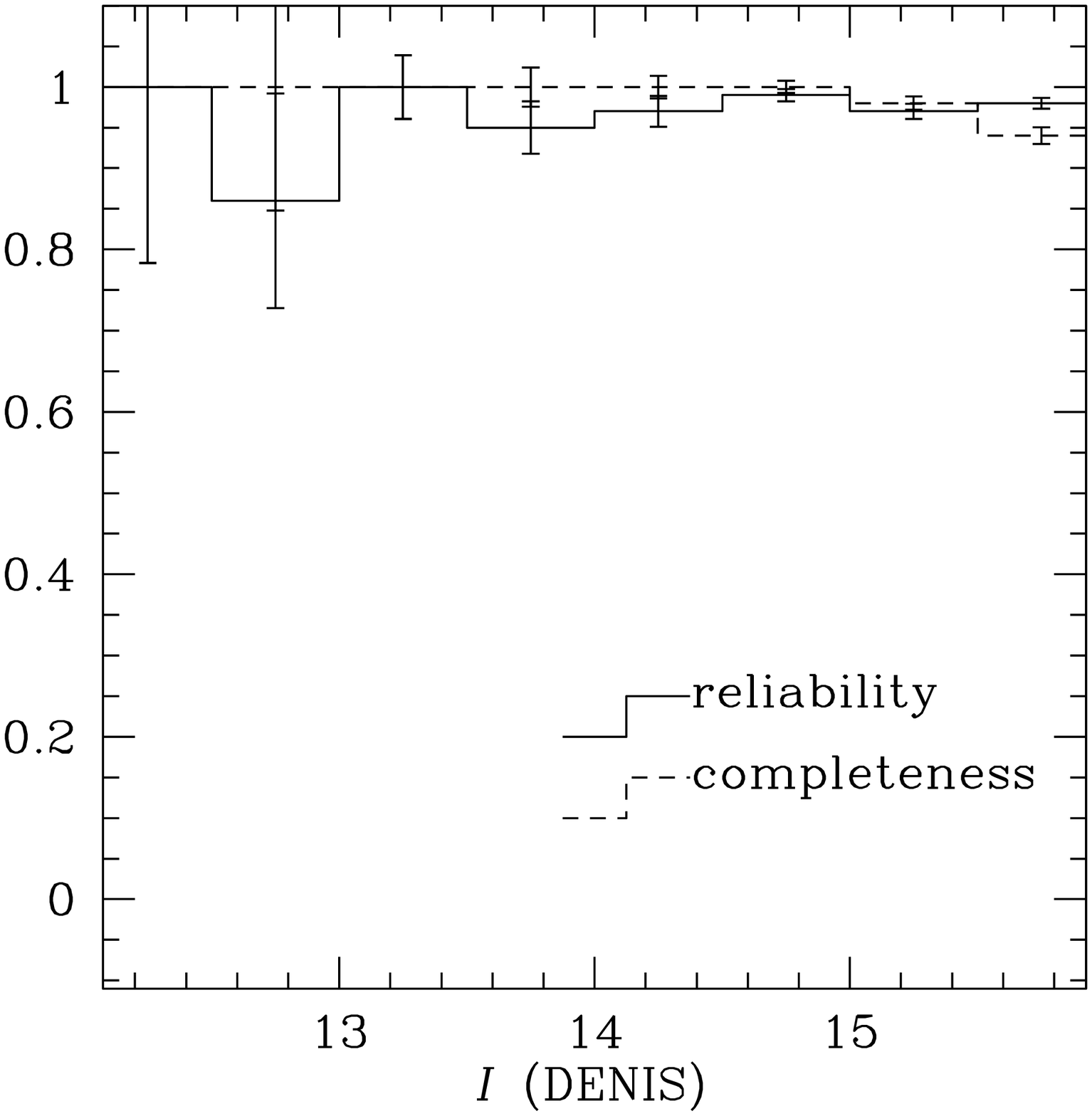}}}
{\resizebox{!}{0.49\hsize}{\includegraphics{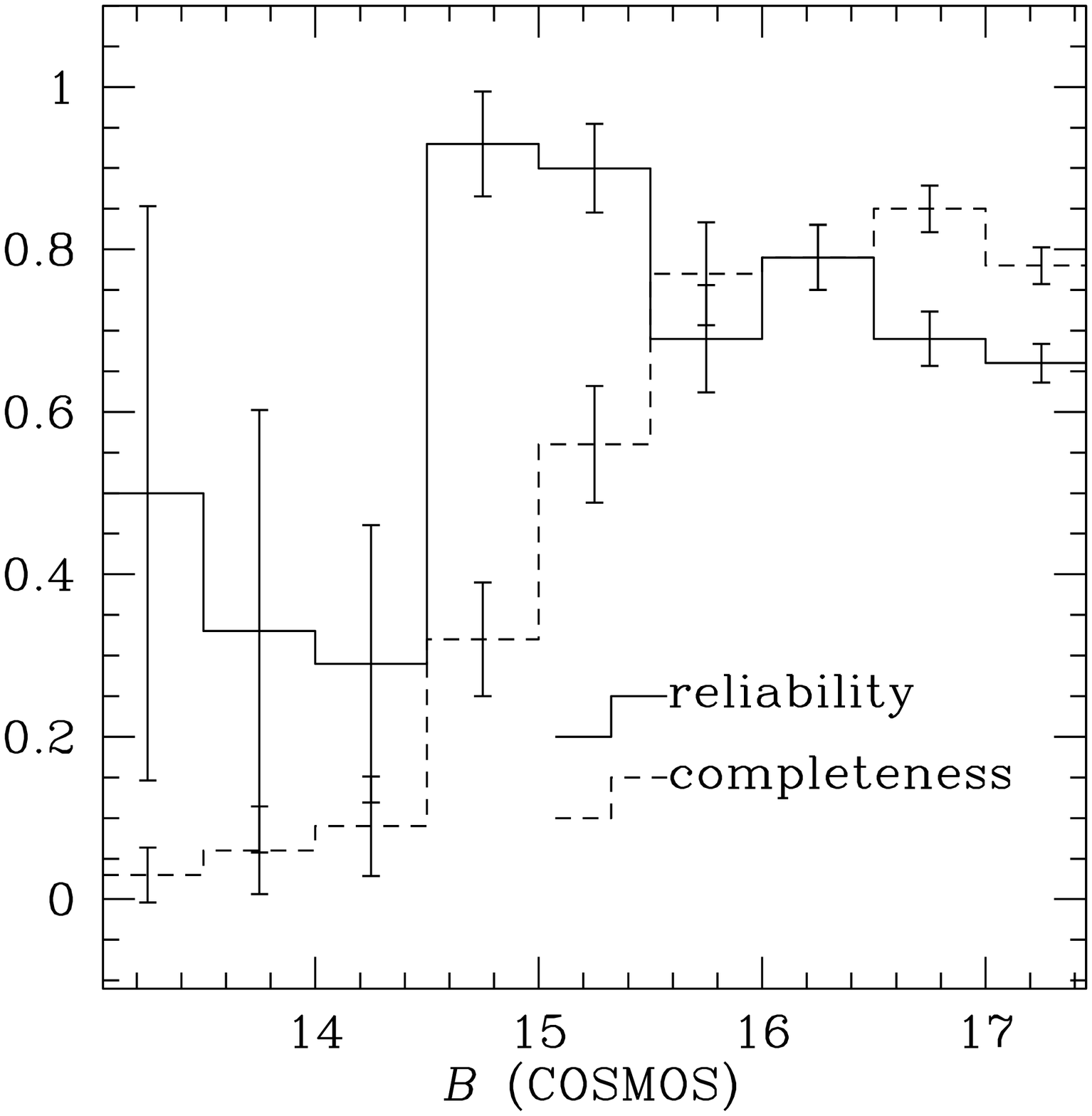}}}
\caption{Completeness ({\it dashed\/}) and reliability ({\it
solid histograms\/}) for {\sf DENIS} ({\it left\/}) and {\sf COSMOS} ({\it
right\/}) 
galaxy extraction.}
\end{center}
\end{figure}

Figure 1 shows the resulting {\sf DENIS} ({\it left\/}) and {\sf COSMOS}
({\it right\/}) 
completeness ({\it dashed\/}) 
and reliability ({\it solid\/}) as a function of magnitude. While {\sf DENIS}
achieves 
better than 95\% completeness and reliability over the full range $13 < I <
16$, {\sf COSMOS} has worse than 80\% completeness and reliability at $B =
17.5$, 
and the numbers get even worse at brighter magnitudes: in particular the
completeness falls to below 10\% at $B < 14.5$.
At bright magnitudes, {\sf COSMOS} suffers from the non-linearity of the plate
photometry. Its reliability suffers from the blending of low non-linear peaks of double stars, while its strong incompleteness is caused by
systematic over-compensation of the photometric non-linearity, thus bright
galaxies are classified as even brighter stars.

\section{The {\sf 6dF} redshift survey}

The strong case for spectroscopic followups of near-IR selected galaxy
samples, coupled with the realization that Schmidt telescopes offer wide
enough fields to sample most efficiently the sky for shallow samples of 5--10
galaxies deg$^{-2}$,
has led the {\sf AAO} to replace its existing manually-configured {\sf FLAIR
II} 
92-fiber system on the {\sf UKST} with a robotized, 150-fiber
system, called {\sf 6dF} 
(for Six Degree 
Field). The {\sf 6dF} instrument, currently under construction uses an
$r-\theta$ positioner, adapted to the curved focal surface of the {\sf UKST}.


The {\sf AAO} has suggested granting 
300 nights of dark/grey time during 2001--2003 
to a Scientific Advisory Group ({\sf 6dFSAG}, see
Acknowledgments, below), 
to conduct a redshift survey based upon near-IR selected galaxy
samples.
The {\sf 6dFSAG} is envisioning a complete spectroscopic followup of $\simeq
120$,$000$ {\sf DENIS}/{\sf 2MASS}
selected southern galaxies.
The primary sample should be a {\sf 2MASS}-$K$ or {\sf DENIS}-$J$
selected sample. 
Moreover, we will add galaxies to form 
complete subsamples in {\sf DENIS} $I$ and {\sf 2MASS} $H$, as well as
{\sf SuperCOSMOS} $B$ and $R$.

Because fibers cannot be placed closer than $5'$ to one another, we envision
an adaptive tiling scheme, with typically 2 exposures per effective
field 
to minimize the 
losses due to crowding. At a galaxy density of 7 deg$^{-2}$, our 27 deg$^2$
field will have an average of 189 galaxies. If 10 fibers are reserved for
sky, we will have $2\times(150-10)=280$
fibers available for objects per field, hence an average of $280-189=91$
spare fibers, which we intend to use for other classes of objects such as
quasars and/or {\sf IRAS} galaxies, as well as allow for cosmic variance of the
field density. 

\section{The {\sf 6dF} and {\sf DENIS-HI} peculiar velocity surveys}
\label{6df4d}

The principal motivation of the {\sf 6dFSAG} is to perform a peculiar velocity
survey in the full ($|b| > 10^\circ$) southern hemisphere.
Assuming 3/4 dark or grey {\sf UKST} 
telescope time and two 4 hours exposures per night,
we expect to measure $1\,\rm \AA$ resolution spectra for
roughly 12,000 ellipticals and lenticulars at $cz < 15$,$000 \, \rm km \,
s^{-1}$. 
To limit the
survey duration, only the densest half of the fields will be observed.
 The line-widths will be coupled
with surface photometry (using {\sf DENIS} $I$ or {\sf 2MASS} $K$ images), to
yield distances 
(with the $D_n\!-\!\sigma$ relation)  and hence peculiar velocities.
We hope to be awarded the time to pursue the peculiar velocity survey during
the period 2003--2005.


In the meanwhile, the {\sf DENIS-HI}  consortium has just begun a
spectroscopic followup of {\sf DENIS} southern galaxies with the Nan\c{c}ay
radio-telescope.
We estimate that given one-sixth the Nan\c{c}ay time, we can reach the
estimated  5000 inclined
spiral galaxies at $-38^\circ < \delta < +2^\circ$ with $I <
14.5$ and $cz < 10$,$000 \, \rm km \, s^{-1}$ in 2000--2004, 
thanks to the 5-fold increase in sensitivity
and velocity coverage that Nan\c{c}ay should obtain in the fall of 1999.
So far, a hundred spectra have been obtained.

\section{Perspectives}

The {\sf DENIS-HI} and {\sf 6dF} peculiar velocity surveys should increase
the coverage of the southern hemisphere from 1500 to 17,000 galaxies, and
within $cz = 6000 \, \rm km \, s^{-1}$, the space density of objects will be
4 times higher than today.
Therefore, the problems related to the 
sparseness of the current peculiar velocity samples should disappear
with these two surveys, leading to more accurate estimates of bulk flow, $\Omega_{\rm matter}^{0.6}/$bias,
$\Omega_{\rm matter}$ itself,  
and especially the primordial density fluctuation spectrum.

\medskip

\acknowledgments

I thank members of the {\sf DENIS} team, especially N. Epchtein,
P. Fouqu\'e, and J. Borsenberger.
Thanks also to F. Giraud who participated in the comparison of {\sf DENIS}
and {\sf COSMOS}, and E. Bertin for useful discussions
on the 
matter. I also acknowledge my colleagues on the {\sf 6dF} Science Advisory
Group: 
M. Colless, J. Huchra,
O. Lahav, J. Lucey, Q. Parker, E. Sadler, W. Saunders, and F. Watson
and on
the {\sf DENIS-HI} team: G. Theureau, I. Vauglin, G. Paturel, L. Bottinelli,
J.-M. Martin, and W. van Driel.

\end{document}